\documentclass[10pt]{article}

\def\CRAS{C.~R.~Acad.~Sc.~Paris}

\def \D {\hbox{d}}
\def\ea{e_1}  %NDLR Choisir une notation
\def\et{e_t}  %NDLR Choisir une notation

\begin{document}

\title{Hamiltonians with two degrees of freedom
admitting a singlevalued general solution\footnote
{Analysis in theory and applications (Nanjing), to appear, 2005.
Nanjing, 20--24 July 2004.
Corresponding author RC. Preprint S2005/006. nlin.SI/0507012}
}

\author{Robert CONTE~$^\dag$,
Micheline MUSETTE~$^\ddag$
and
Caroline VERHOEVEN~$^\ddag$
{}
\\ \dag Service de physique de l'\'etat condens\'e (URA 2464), CEA--Saclay
\\ F--91191 Gif-sur-Yvette Cedex, France
%\ E-mail:  \begin{verbatim} Conte@drecam.saclay.cea.fr \end{verbatim}
\\ E-mail:  Conte@drecam.saclay.cea.fr
{}\\
\\ \ddag Dienst Theoretische Natuurkunde, Vrije Universiteit Brussel
and
\\~~International Solvay Institutes for Physics and Chemistry
\\~~Pleinlaan 2, B--1050 Brussels, Belgium
\\~~E-mail: MMusette@vub.ac.be, CVerhoev@vub.ac.be
}

\maketitle

\hfill % \today

{\vglue -10.0 truemm}
{\vskip -10.0 truemm}

\begin{abstract}
\noindent
Following the basic principles stated by Painlev\'e,
we first revisit the process of selecting the admissible
time-independent Hamiltonians $H=(p_1^2+p_2^2)/2+V(q_1,q_2)$
whose some integer power $q_j^{n_j}(t)$ of the general solution
is a singlevalued function of the complex time $t$.
In addition to the well known rational potentials $V$ of H\'enon-Heiles,
this selects possible cases with a trigonometric dependence of $V$ on $q_j$.
Then, by establishing the relevant confluences,
we restrict the question of the explicit integration of the seven
(three ``cubic'' plus four ``quartic'') rational H\'enon-Heiles cases
to the quartic cases.
Finally, we perform the explicit integration of the quartic cases,
thus proving that the seven rational cases have a meromorphic general solution
explicitly given by a genus two hyperelliptic function.
\end{abstract}

%\noindent \textit{Running title}:


%\noindent \textit{MSC 2000}

\noindent \textit{PACS 1995}:
02.30.Hq, % Ordinary differential equations
03.40     % Classical mechanics

\baselineskip=14truept % 68 pages

% FOR DOUBLE-SPACING, REMOVE THE TWO FOLLOWING %
%\baselineskip=24truept
%\renewcommand{\baselinestretch}{2.0}

\tableofcontents

%\vfill \eject

% ==========================================================================
\section{Introduction}
\indent

We consider the most general
two-degree of freedom, classical, time-independent Hamiltonian
of the physical type
(i.e.~the sum of a kinetic energy and a potential energy),
\begin{eqnarray}
H
& =&
 \frac{1}{2} (p_1^2 + p_2^2) + V(q_1,q_2),
\label{eqH2TV}
\end{eqnarray}
and the problem which we address is
to determine all the potentials $V(q_1,q_2)$
such that some unspecified integer power $q_j^{n_j}(t)$ of the
general solution is a single valued function of the complex time $t$.

In the case of one degree of freedom,
this problem only admits two solutions,
% Briot et Bouquet (ordre 1)
\begin{eqnarray}
n=\pm 1,\
H
& =&
 \frac{p^2}{2} + \sum_{j=1}^{4} c_j q^j,\
\\
n=\pm 2,\
H
& =&
 \frac{p^2}{2} + a q^{-2} + c_2 q^2 + c_4 q^4 + c_6 q^6.
\end{eqnarray}
In both cases, $q^n$ is an elliptic function
and, in the second case, $q$ is generically multivalued.

This property that the general solution of a differential equation
is singlevalued,
except maybe at the singularities of the equation itself,
is called the \textit{Painlev\'e property} (PP)
\cite{Cargese1996Conte}.

The equations of motion for $q_j(t)$ are obtained by eliminating
the momenta $p_1,p_2$ between the Hamilton's equations of motion,
\begin{eqnarray}
& &
\frac{\D p_j}{\D t}=-\frac{\partial V(q_1,q_2)}{\partial q_j},\
\frac{\D q_j}{\D t}=q_j'=p_j,\
j=1,2
\label{eqHamiltonEquations}
\end{eqnarray}
which results into the system of two coupled second order
ordinary differential equations (ODE)
\begin{eqnarray}
& &
q_j''+\frac{\partial V(q_1,q_2)}{\partial q_j}=0,\
j=1,2
\label{eqHamiltonianSystemOrder4}
\end{eqnarray}
together with the first integral
\begin{eqnarray}
H \equiv \frac{{q_1'}^2}{2} + \frac{{q_2'}^2}{2} + V(q_1,q_2)=E.
\label{eqHamiltonianFirstIntegral}
\end{eqnarray}

To prove the Painlev\'e property, one must perform the two following steps.
\begin{enumerate}
\item
Generation of necessary conditions for the single valuedness
of the general solution.
This step is algorithmic and known as the \textit{Painlev\'e test}
\cite{Cargese1996Conte}.
However, its output is only a set of necessary conditions,
in our case a selection of candidate potentials $V(q_1,q_2)$.

\item
For each such candidate $V$, explicit integration of the equations of
motion,
so as to indeed check the single valuedness of the general solution.
\end{enumerate}

In section \ref{sectionSelection},
we select the potentials $V(q_1,q_2)$ according to the prescriptions
of the Painlev\'e test.
In section \ref{sectionThe_seven},
we present the seven so-called H\'enon-Heiles Hamiltonians.
In section \ref{sectionConfluences},
we establish confluences from one subset of these seven Hamiltonians
to another subset,
thus restricting the question of their explicit integration to the first
subset.
In sections \ref{sectionPoint} and \ref{sectionBirational},
we recall the explicit integration of this first subset,
the so-called ``quartic'' cases.

% ==========================================================================
\section{Selection of the candidate potentials $V$}
\label{sectionSelection}
\indent

The difficulty is that very few results exist concerning coupled systems
of nonlinear ODEs possessing the Painlev\'e property.
On the contrary, for a single ODE,
many results exist,
either as exhaustive lists of equations in a given class
(e.g.~second order first degree)
which possess the PP,
or as precise necessary conditions to be satisfied.

Let us therefore build, by elimination of $q_2$,
a single ODE in $q_1(t)$ in a class at least partially studied.
Taking the shorthand notation
\begin{eqnarray}
& &
V_{m,n}=\frac{\partial^{m+n} V(q_1,q_2)}{\partial q_1^m \partial q_2^n},
%\label{eqNotationVmn}
\end{eqnarray}
one eliminates $q_1'',q_2'',q_2'$ between the system of four equations
made of (\ref{eqHamiltonianSystemOrder4}) and the first two derivatives of
\begin{eqnarray}
& &
q_1''+V_{10}(q_1,q_2)=0.
\label{eqHamiltonianSystem_q1second}
\end{eqnarray}
This yields
\begin{eqnarray}
& & \left\lbrace
\begin{array}{ll}
\displaystyle{
q_1''=-V_{10}(q_1,q_2),
}
\\
\displaystyle{
q_2''=-V_{01}(q_1,q_2),
}
\\
\displaystyle{
q_2'=-\frac{q_1'''}{V_{11}} - \frac{V_{20}}{V_{11}} q_1',
}
\end{array}
\right.
\end{eqnarray}
and the fourth order first degree ODE for $q_1(t)$
\begin{eqnarray}
& &
-q_1''''
 - V_{12} \left(\frac{q_1'''}{V_{11}} + \frac{V_{20}}{V_{11}} q_1' \right)^2
 +2V_{21} \left(\frac{q_1'''}{V_{11}} + \frac{V_{20}}{V_{11}} q_1' \right)
    q_1'
 -V_{20} q_1'' -V_{30} {q_1'}^2 + V_{01} V_{11}=0,
\label{eq_q1_Order4_bad}
\end{eqnarray}
in which the coefficients $V_{mn}$ only depend on $(q_1'',q_1)$
after the (implicit) elimination of $q_2$ from
(\ref{eqHamiltonianSystem_q1second}).

The similar elimination with (\ref{eqHamiltonianFirstIntegral}) yields
the third order second degree ODE
\begin{eqnarray}
& &
\frac{{q_1'}^2}{2}
 + \frac{1}{2}
  \left(\frac{q_1'''}{V_{11}} + \frac{V_{20}}{V_{11}} q_1' \right)^2
 + V - E=0.
\label{eq_q1_Order3}
\end{eqnarray}

None of the two ODEs
(\ref{eq_q1_Order4_bad}), (\ref{eq_q1_Order3})
is very helpful to generate necessary conditions on $V$,
but the combination which eliminates ${q_1'''}^2$,
namely
\begin{eqnarray}
& &
-q_1''''
+2 \frac{V_{21}}{V_{11}} q_1' q_1'''
-V_{20} q_1''
+\left(2 \frac{V_{21} V_{20}}{V_{11}}-V_{30}+V_{12}\right) {q_1'}^2
+V_{01} V_{11}
+ 2 (V-E) V_{12}
=0,
\label{eq_q1_Order4_good}
\end{eqnarray}
is quite helpful since it has only degree one in $q_1'''$.

Indeed,
in 1902 Painlev\'e \cite[p.~74]{PaiActa} established necessary conditions
for an $n$-th order first degree ODE
\begin{eqnarray}
& &
u^{(n)}=F(u^{(n-1)},\dots,u,t),
\label{eqOrdern}
\end{eqnarray}
to possess the PP,
when $F$ is assumed rational in $u^{(n-1)},u^{(n-2)}$,
algebraic in $u^{(n-3)},\dots,u$, and analytic in $t$
(we will also assume such a dependence for (\ref{eq_q1_Order4_good})).

The \textit{first necessary condition} is that the highest derivative
$u^{(n)}$, as a function of the next highest derivative $u^{(n-1)}$,
be a polynomial of degree at most two
(i.e.~that the ODE for $u^{(n-1)}$ be of Riccati type),
\begin{eqnarray}
& &
u^{(n)}=\sum_{j=0}^2 A_j(u^{(n-2)},\dots,u,t) \left(u^{(n-1)}\right)^j,
\end{eqnarray}
which is indeed the case for both
(\ref{eq_q1_Order4_bad})
and
(\ref{eq_q1_Order4_good}).

The \textit{second necessary condition} states that,
as a function of the second next highest derivative $u^{(n-2)}$,
each coefficient $A_j$ has for only singularities simple poles,
the poles of $A_1$ and $A_0$ are among those of $A_2$,
and the difference between the degrees of the numerator and denominator
of $A_j$ does not exceed $-1,1,3$ for, respectively, $j=2,1,0$.
When applied to (\ref{eq_q1_Order4_good}),
since $A_2$ is identically zero and thus has no poles
(this feature is precisely the advantage of
(\ref{eq_q1_Order4_good}) over (\ref{eq_q1_Order4_bad})),
this latter condition
requires that the coefficients
$A_1 \equiv 2 (V_{21}/V_{11}) q_1'$
and $A_0$
(in which, as always,
$q_2$ is eliminated from (\ref{eqHamiltonianSystem_q1second}))
be polynomials in $q_1''$ with maximal respective degrees $1$ and $3$.
The necessary condition arising from $A_1$ is
\begin{eqnarray}
& &
\exists F_1,G_1:\ \frac{V_{21}}{V_{11}}=F_1(q_1) q_1'' + G_1(q_1).
\label{eqPDEV}
\end{eqnarray}
Assuming the additional condition $F_1(q_1)=0$,
the partial differential equation (\ref{eqPDEV}) is integrated as
\begin{eqnarray}
& &
V(q_1,q_2)=f_1(q_1) f_2(q_2) + h_1(q_1) + h_2(q_2),\
f_1' f_2' \not=0,
\label{eqV}
\end{eqnarray}
in which the four functions must be further constrained.

Instead of $f_2(q_2)$,
let us introduce its inverse function $F_2(r_1)$
from (\ref{eqHamiltonianSystem_q1second}),
\begin{eqnarray}
& &
r_1=f_2(q_2),\
q_2=F_2(r_1),\
r_1=-\frac{q_1''-h_1'(q_1)}{f_1'(q_1)},
\label{eq_q2_functionof_q1second}
\end{eqnarray}
which implies
\begin{eqnarray}
& &
f_2'(q_2)=\frac{1}{F_2'(r_1)},\
f_2''(q_2)=-\frac{F_2''(r_1)}{\left(F_2'(r_1)\right)^3}.
\label{eq_g2der1der2}
\end{eqnarray}
The equation (\ref{eq_q1_Order4_good}) then becomes
\begin{eqnarray}
& &
-q_1''''
+ \frac{f_1''}{f_1'} \left(2 q_1' q_1''' +{q_1''}^2\right)
+\left(2 \frac{f_1'''}{f_1'}-\frac{{f_1''}^2}{{f_1'}^2}\right) {q_1'}^2 q_1''
\nonumber \\ & & \phantom{1234}
+\left(\frac{f_1'' h_1'}{f_1'}- h_1'' \right) q_1''
+\left(\frac{f_1'''}{f_1'} h_1' +2 \frac{f_1''}{f_1'} h_1'' - h_1'''
      - 2 \frac{{f_1''}^2}{{f_1'}^2} h_1'
\right) {q_1'}^2
\nonumber \\ & & \phantom{1234}
-\left(f_1' {q_1'}^2 - 2 f_1 q_1'' -2 E f_1'+ 2 (h_1 f_1'-h_1' f_1 )\right)
\frac{F_2''(r_1)}{\left(F_2'(r_1)\right)^3}
\nonumber \\ & & \phantom{1234}
+ \frac{f_1 f_1'}{\left(F_2'(r_1)\right)^2}
+ f_1' \frac{\D}{\D r_1} \left(\left(F_2'(r_1)\right)^{-2} h_2\right)
=0,
\label{eq_q1_Order4_form2}
\end{eqnarray}
in which the dependence on $q_1''$ is also
implicit through the dependence on $r_1$ in the last two lines,
and $A_0$ must be a polynomial in $q_1''$ of degree at most three.

The term ${q_1'}^2$ first constrains $F_2$,
\begin{eqnarray}
& &
\left(f_2'(q_2)\right)^2=\frac{1}{\left(F_2'(r_1)\right)^2}=P_4(r_1),\
f_2''(q_2)=-\frac{F_2''(r_1)}{\left(F_2'(r_1)\right)^3}=\frac{1}{2} P_4'(r_1),\
P_4(r_1)=\sum_{j=0}^4 d_j r_1^j,
\end{eqnarray}
in which the coefficients $d_j$ are constant,
then the term depending on $h_2$ generates the constraint
\begin{eqnarray}
& &
h_2=\frac{Q_5(r_1)}{P_4(r_1)},\
Q_5(r_1)=\sum_{j=0}^5 e_j r_1^j,
\label{eqh2ofr1}
\end{eqnarray}
in which the coefficients $e_j$ are constant.
The resulting fourth power of $q_1''$,
\begin{eqnarray}
& &
-5 \frac{e_5 + d_4 f_1}{{f_1'}^3} {q_1''}^4
\end{eqnarray}
must be canceled, which implies $d_4=0$ and $e_5=0$.
Finally, if one performs the $\alpha$-transformation
\begin{eqnarray}
& &
(t,q_1) \to (T,Q_1):\
t=\varepsilon T,\
q_1=a + \varepsilon Q_1,
\label{eqTransfo_alpha}
\end{eqnarray}
the limit $\varepsilon \to 0$ of (\ref{eq_q1_Order4_form2}),
\begin{eqnarray}
& &
-Q_1''''
+ 4 \frac{e_4 + d_3 f_1(a) }{\left(f_1'(a)\right)^2} {Q_1''}^3=0,
\end{eqnarray}
must have the PP, which requires $d_3=0$ and $e_4=0$ since the constant
$a$ is arbitrary.

The other $\alpha$-transformation
\begin{eqnarray}
& &
(t,q_1) \to (T,Q_1):\
t=\varepsilon T,\
q_1=Q_1,
\\
& &
-Q_1''''
+ 2 \frac{f_1''}{f_1'} Q_1' Q_1'''
+ \left(\frac{f_1''}{f_1'}+3 d_2 \frac{f_1'}{f_1}+3 \frac{e_3}{f_1'}\right)
  {Q_1''}^2
+ \left(\frac{f_1'''}{f_1'}-2 \frac{{f_1''}^2}{{f_1'}^2} - d_2\right)
   {Q'}^2 Q_1''
=0,
\end{eqnarray}
should constrain $f_1(q_1)$,
but we have not further explored this way.

The differential equation obeyed by $f_2(q_2)$,
\begin{eqnarray}
& &
\left(\frac{\D f_2}{\D q_2}\right)^2
=d_0+d_1 f_2 +d_2 f_2^2,
\label{eqODE_g2_of_q2}
\end{eqnarray}
has the Painlev\'e property,
and its solutions are displayed in Table \ref{Table1}.
\tabcolsep=1.5truemm
\tabcolsep=0.5truemm

\begin{table}[h] % [p]
\caption[garbage]{
Admissible potentials $V(q_1,q_2)$
selected by the condition of singlevaluedness of $q_1(t)$.
The potential must have the form (\ref{eqV}),
in which $f_2$ and $h_2$ have only four admissible values.
}
\vspace{0.2truecm}
\begin{center}
\begin{tabular}{| l | l | l |}
\hline % \hline % ********************************************************
$(d_2   ,d_1   ,d_0   )$ & $f_2$               & Terms in $h_2$
\\ \hline % \hline % ********************************************************
$(\not=0,     0,\not=0)$ & $a_2 \sinh b_2 q_2$ & \dots
\\ \hline % \hline % ********************************************************
$(\not=0,     0,     0)$ & $a_2 e^{b_2 q_2}  $ & \dots
\\ \hline % \hline % ********************************************************
$(     0,\not=0,     0)$ & $a_2 q_2^2 + c_2  $ & $q_2^{-2},q_2^2,q_2^4$
\\ \hline % \hline % ********************************************************
$(     0,     0,\not=0)$ & $b_2 q_2          $ & $q_2^{1},q_2^2,q_2^3$
\\ \hline % \hline % ********************************************************
\end{tabular}
\end{center}
\label{Table1}
\end{table}

\label{Table2} % NDLR To be established soon.

The same study about the single valuedness of $q_1^2$
would lead to another, similar table listing a finite number of
admissible potentials depending on a finite number of arbitrary constants.
This Table \ref{Table2} will be established in a very near future.

To conclude this first part of the Painlev\'e test,
the admissible potentials $V$ for which the general solution
$(q_1^{n_1},q_2^{n_2})$, $n_j=\pm 1$ or $\pm 2$, may be single valued
are built by taking the appropriate information on $(f_j,h_j)$
from Tables \ref{Table1} and \ref{Table2}.
These potentials depend on a finite number of arbitrary constants.

The second part of the Painlev\'e test is very well known
\cite{Kowa1890a,GambierThese} \cite[\S 6.6]{Cargese1996Conte}
and consists in analyzing
the system of two coupled second order ODEs
(\ref{eqHamiltonianSystemOrder4}),
%which is now rational in $q_j$ and its derivatives,
in order to enforce the absence of any branch point
(either algebraic or logarithmic)
whose location depends on the initial conditions (one says \textit{movable}).
This test is well defined only when the ODEs are algebraic,
which is the probable reason for the usual discarding of the
trigonometric cases.
Although the test seems to have never been applied yet to the full rational
cases isolated above,
we will not perform here these lengthy calculations
and directly skip to the question of the explicit integration of
the candidate cases.

% ============================================================================
\section     {The seven H\'enon-Heiles Hamiltonians}
\label{sectionThe_seven}
\indent

Among the set of rational potentials $V(q_1,q_2)$ selected
in section \ref{sectionSelection},
there exists a subset,
called for historical reasons \textit{H\'enon-Heiles Hamiltonians}
\cite{HH}.
These seven potentials,
usually denoted
``cubic'' or ``quartic'' according to their global degree in $(q_1,q_2)$,
were in fact isolated by the condition
that a second integral of the motion should exist \cite{H1987}
(Liouville integrability).
The difference between the two approaches is quite important:
requiring singlevaluedness generates \textit{necessary} conditions
on $V(q_1,q_2)$,
while requiring the existence of a second integral of motion
only results in \textit{sufficient} conditions since $V(q_1,q_2)$ must be
an input.

The cubic case basically arises from $f_1=b_1 q_1,f_2=a_2 q_2^2+ c_2$,
and the quartic case from $f_1=a_1 q_1^2 + c_1,f_2=a_2 q_2^2+ c_2$,
and their usual notation is as follows.

\begin{enumerate}
\item
In the cubic case HH3
\cite{CTW,Fordy1991,CFP1993},
       \index{H\'enon-Heiles Hamiltonian!cubic}
\begin{eqnarray}
H
& =&
 \frac{1}{2} (p_1^2 + p_2^2 + \omega_1 q_1^2 + \omega_2 q_2^2)
    + \alpha q_1 q_2^2 - \frac{\beta}{3} q_1^3
    + \frac{\gamma}{2} q_2^{-2},\
%    + \frac{1}{6} c_4 q_2^{-6}
\alpha \not=0
\label{eqHH0}
%\\
%& & q_1'' + \omega_1 q_1 - \beta q_1^2 + \alpha q_2^2 = 0,
%\label{eqHH1}
%\\
%& & q_2'' + \omega_2 q_2 + 2 \alpha q_1 q_2 - \gamma q_2^{-3}
%% - c_4 q_2^{-7}
%=0,
%\label{eqHH2}
\end{eqnarray}
in which the constants $\alpha,\beta,\omega_1,\omega_2$ and $\gamma$
can only take the three sets of values,
\begin{eqnarray}
\hbox{(SK)} : & & \beta/ \alpha=-1, \omega_1=\omega_2,\
\label{eqHH3SKcond} \\
\hbox{(KdV5)} : & & \beta/ \alpha=-6,\
\label{eqHH3K5cond}\\
\hbox{(KK)} : & & \beta/ \alpha=-16, \omega_1=16 \omega_2.
\label{eqHH3KKcond}
\end{eqnarray}
\item
In the quartic case HH4
\cite{RDG1982,GDR1983},
\index{H\'enon-Heiles Hamiltonian!quartic}
\begin{eqnarray}
H & = &
\frac{1}{2}(P_1^2+P_2^2+\Omega_1 Q_1^2+\Omega_2 Q_2^2)
 +C Q_1^4+ B Q_1^2 Q_2^2 + A Q_2^4
\nonumber
\\
& &
 +\frac{1}{2}\left(\frac{\alpha}{Q_1^2}+\frac{\beta}{Q_2^2}\right)
 + \gamma Q_1,\ B \not=0,
\label{eqHH40}
%\\
%& & Q_1''+\Omega_1 Q_1 + 4 C Q_1^3 + 2 B Q_1 Q_2^2 - \alpha Q_1^{-3} + \gamma=0,
%\label{eqHH41}
%\\
%& & Q_2''+\Omega_2 Q_2 + 4 A Q_2^3 + 2 B Q_2 Q_1^2 - \beta  Q_2^{-3}=0,
%\label{eqHH42}
\end{eqnarray}
in which the constants
$A,B,C,\alpha,\beta,\gamma,\Omega_1$ and $\Omega_2$
can only take the four values
(the notation $A:B:C=p:q:r$ stands for $A/p=B/q=C/r=\hbox{arbitrary}$),
\begin{eqnarray}
& & \left\lbrace
\begin{array}{ll}
\displaystyle{
%\hbox{(NLS)} :
A:B:C=1:2:1,\ \gamma=0,
}
\\
\displaystyle{
%\hbox{(KP-1)} :
A:B:C=1:6:1,\ \gamma=0,\ \Omega_1=\Omega_2,
}
\\
\displaystyle{
%\hbox{(KP-1)} :
A:B:C=1:6:8,\ \alpha=0,\ \Omega_1=4\Omega_2,
}
\\
\displaystyle{
%\hbox{(B-KP,C-KP)} :
A:B:C=1:12:16,\ \gamma=0,\ \Omega_1=4\Omega_2.
}
\end{array}
\right.
\label{eqHH4NLScond}
\end{eqnarray}
\end{enumerate}

For each of the seven cases so isolated
there exists a second constant of the motion $K$
\cite{Drach1919KdV,BEF1995b,H1984} % cubic
\cite{H1987,BakerThesis,BEF1995b} % quartic
in involution with the Hamiltonian,
\begin{eqnarray}
{\hskip -00.0truemm}
\hbox{(SK)} : K & = &
\left(3 p_1 p_2 + \alpha q_2 (3 q_1^2 + q_2^2) + 3 \omega_2 q_1 q_2 \right)^2
+ 3 \gamma (3 p_1^2 q_2^{-2} + 4 \alpha q_1 + 2 \omega_2),\
\label{eqHH3SKSecond}
\\
{\hskip -00.0truemm}
\hbox{(KdV5)} : K & = &
4 \alpha p_2 (q_2 p_1 - q_1 p_2)
+ (4 \omega_2 - \omega_1) (p_2^2 + \omega_2 q_2^2 + \gamma q_2^{-2})
\nonumber \\ & &
+ \alpha^2 q_2^2 (4 q_1^2 + q_2^2)
+ 4 \alpha q_1 (\omega_2 q_2^2 - \gamma q_2^{-2}),
\label{eqHH3K5Second}
\\
{\hskip -00.0truemm}
\hbox{(KK)} : K & = &
(3 p_2^2 + 3 \omega_2 q_2^2 + 3 \gamma q_2^{-2})^2
+ 12 \alpha p_2 q_2^2 (3 q_1 p_2 - q_2 p_1)
\nonumber \\ & &
- 2 \alpha^2 q_2^4 (6 q_1^2 + q_2^2)
+ 12 \alpha q_1 (-\omega_2 q_2^4 + \gamma)
- 12 \omega_2 \gamma,
\label{eqHH3KKSecond}
%\\
%{\hskip -00.0truemm}
%\hbox{quartic} : K & = & \hbox{see }
%(\ref{eqHH401:2:1K}), (\ref{eqHH40161}), (\ref{eqHH40168}), (\ref{eqHH401:12:16K}).
\\
{\hskip -00.0truemm}
\hbox{1:2:1} : K & = &
\left(Q_2 P_1 - Q_1 P_2 \right)^2
+ Q_2^2 \frac{\alpha}{Q_1^2} + Q_1^2 \frac{\beta}{Q_2^2}
\nonumber
\\
& &
-\frac{\Omega_1-\Omega_2}{2}
\left(P_1^2-P_2^2+Q_1^4-Q_2^4+\Omega_1 Q_1^2 - \Omega_2 Q_2^2
+ \frac{\alpha}{Q_1^2} - \frac{\beta}{Q_2^2}
 \right),\
 A=\frac{1}{2},
\label{eqHH401:2:1K}
\\
{\hskip -00.0truemm}
\hbox{1:6:1} : K & = &
\left(
P_1 P_2 + Q_1 Q_2 \left(-\frac{Q_1^2+Q_2^2}{8}+\Omega_1 \right)
\right)^2
\\
& &
%\phantom{1234}
- P_2^2 \frac{\kappa_1^2}{Q_1^2}
- P_1^2 \frac{\kappa_2^2}{Q_2^2}
+\frac{1}{4}\left(\kappa_1^2 Q_2^2 + \kappa_2^2 Q_1^2 \right)
+\frac{\kappa_1^2 \kappa_2^2}{Q_1^2 Q_2^2},\
\alpha=-\kappa_1^2,\
\beta= - \kappa_2^2,\
A=-\frac{1}{32},
\\
{\hskip -00.0truemm}
\hbox{1:6:8} : K & = &
\left(
P_2^2-\frac{Q_2^2}{16}(2 Q_2^2+4 Q_1^2+\Omega_2)
     +\frac{\beta}{Q_2^2}
\right)^2
-\frac{1}{4}Q_2^2(Q_2 P_1 - 2 Q_1 P_2)^2
\\
& &
%\phantom{1234}
+\gamma
\left(
-2 \gamma Q_2^2
-4 Q_2 P_1 P_2
+\frac{1}{2} Q_1 Q_2^4
+ Q_1^3 Q_2^2
+4 Q_1 P_2^2
-4 \Omega_2 Q_1 Q_2^2
+ 4 Q_1 \frac{\beta}{Q_2^2}
\right),\
\\
& &
%\phantom{1234}
A=-\frac{1}{16},
\\
{\hskip -00.0truemm}
\hbox{1:12:16} : K & = &
\left(8 (Q_2 P_1 - Q_1 P_2) P_2 - Q_1 Q_2^4 - 2 Q_1^3 Q_2^2
 + 2 \Omega_1 Q_1 Q_2^2 - 8 Q_1 \frac{\beta}{Q_2^2} \right)^2
\\
& &
%\phantom{1234}
+\frac{32 \alpha}{5}
\left(Q_2^4 + 10 \frac{Q_2^2 P_2^2}{Q_1^2}\right),\
A=-\frac{1}{32}.
\end{eqnarray}

% ============================================================================
\section     {Confluences from HH4 to HH3}
\label{sectionConfluences}
\indent

As is well known,
there exists a limiting process (\textit{confluence})
which,
starting from the
Gauss hypergeometric equation,
generates the sequence:
Whittaker equation,
Hermite-Weber and Bessel equations,
Airy equation.
The general solution of all these equations is therefore deductible
from that of the Gauss hypergeometric equation.

A similar confluence also exists
\cite{Rom1995b,V2003,CMVAngers2004}
among the seven HH Hamiltonians,
and each cubic case can be obtained
by a confluence of at least one quartic case.

The following confluences have been established,
\begin{eqnarray}
& &
\left\lbrace
\begin{array}{ll}
\displaystyle{
\hbox{HH4 1:2:1 } \to \hbox{ HH3 KdV5},\
}
\\
\displaystyle{
\hbox{HH4 1:6:8 }
%  +\frac{\alpha}{2 q_1^2}+\frac{\beta}{2 q_2^2}+\frac{\nu}{q_2^6}
 \to \hbox{ HH3 KK},\
}
\\
\displaystyle{
\hbox{HH4 1:6:8 }
%  +\frac{\alpha}{2 q_1^2}+\frac{\beta}{2 q_2^2}+\frac{\nu}{q_2^6}
 \to \hbox{ HH3 KdV5},\
}
\\
\displaystyle{
\hbox{HH4 1:12:16 } \to \hbox{ HH3 KK}.
}
\\
\displaystyle{
\hbox{HH4 1:12:16 } \to \hbox{ HH3 SK}.
}
\end{array}
\right.
\end{eqnarray}
The absence of any confluence originating from HH4 1:6:1
still has to be explained.

Consider for instance the quartic 1:12:16,
\begin{eqnarray}
& &
\left\lbrace
\begin{array}{ll}
\displaystyle{
h_{1:12:16}(t) =
\frac{1}{2}(p_1^2+p_2^2) + \frac{\omega}{8} (4 q_1^2+ q_2^2)
 - \frac{n}{32} (16 q_1^4+ 12 q_1^2 q_2^2 + q_2^4)
}
\\
\displaystyle{
\phantom{1234}
 +\frac{1}{2}\left(\frac{\alpha}{q_1^2}+\frac{\beta}{q_2^2}\right).
}
\end{array}
\right.
\end{eqnarray}
It admits a confluence to both the HH3 KK and SK cases,
\begin{eqnarray}
& &
\left\lbrace
\begin{array}{ll}
\displaystyle{
H_{\rm KK}(T)=
 \frac{1}{2} (P_1^2 + P_2^2) + \frac{\Omega}{2} (16 Q_1^2 + Q_2^2)
    + N \left(Q_1 Q_2^2 + \frac{16}{3} Q_1^3\right)
     + \frac{B}{2 Q_2^2},
}
\\
\displaystyle{
H_{\rm SK}(T)=
 \frac{1}{2} (P_1^2 + P_2^2) + \frac{\Omega}{2} (Q_1^2 + Q_2^2)
    + N \left(Q_1 Q_2^2 + \frac{1}{3} Q_1^3\right)
     + \frac{B}{2 Q_2^2},
}
\end{array}
\right.
\end{eqnarray}
they are (the integers $\et$ and $\ea$ can be chosen arbitrarily),
\begin{eqnarray}
& &
\hbox{1:12:16 }\to\hbox{ KK}
\left\lbrace
\begin{array}{ll}
\displaystyle{
t=\varepsilon^{\et} T,\
q_1=\varepsilon^{\ea} \left(1 + \varepsilon Q_1\right),\
q_2=\varepsilon^{1+\ea} Q_2,\
n=-\frac{4}{3}\varepsilon^{-2 \ea-2 \et -1} N,\
}
\\
\displaystyle{
\alpha=\varepsilon^{4 \ea-2 \et-1}
 \left(-\frac{4}{3} N + 4 \Omega \varepsilon \right),\
\beta=\varepsilon^{4 \ea-2 \et+4} B,\
}
\\
\displaystyle{
\omega=\varepsilon^{-2 \et-1} \left(- 4 N + 4 \Omega \varepsilon \right),\
h=\varepsilon^{2 \ea -2 \et -1}
 \left(- 2 N + 4 \Omega \varepsilon + H \varepsilon^3\right),\
 \varepsilon \to 0,
}
\end{array}
\right.
\end{eqnarray}
and
\begin{eqnarray}
& &
\hbox{1:12:16 }\to\hbox{ SK}
\left\lbrace
\begin{array}{ll}
\displaystyle{
t=\varepsilon^{\et} T,\
q_1=\varepsilon^{\ea} Q_2,\
q_2=\varepsilon^{\ea-1} \left(1 + \varepsilon Q_1\right),\
n=\varepsilon^{-2 \ea-2 \et -1} N,\
}
\\
\displaystyle{
\alpha=\varepsilon^{4 \ea-2 \et} B,\
\beta=\varepsilon^{4 \ea-2 \et-5} \left(\frac{N}{16}
  +\frac{1}{4} \Omega \varepsilon \right),\
}
\\
\displaystyle{
\omega=\varepsilon^{-2 \et-1} \left(\frac{3}{4} N + \Omega \varepsilon \right),\
h=\varepsilon^{2 \ea -2 \et -3}
 \left(\frac{3}{32} N+\frac{1}{4}\Omega \varepsilon + H \varepsilon^3\right).
% \varepsilon \to 0.
}
\end{array}
\right.
\end{eqnarray}
One checks the loss of one parameter in the process,
since the three quartic parameters $(\alpha,\beta,\omega)$ coalesce to
only two cubic parameters $(B,\Omega)$.

{}From the quartic case HH4 1:2:1 to the cubic case HH3 KdV5,
the confluence is
\begin{eqnarray}
& &
\hbox{1:2:1 }\to\hbox{ KdV5}
\left\lbrace
\begin{array}{ll}
\displaystyle{
h_{1:2:1}(t) =
\frac{1}{2}(p_1^2+p_2^2)+\frac{\omega_1}{2} q_1^2+\frac{\omega_2}{2} q_2^2
 - \frac{n}{2} (q_1^4+ 2 q_1^2 q_2^2 + q_2^4)
 +\frac{1}{2}\left(\frac{\alpha}{q_1^2}+\frac{\beta}{q_2^2}\right),
}
\\
\displaystyle{
H_{\rm KdV5}(T)=
 \frac{1}{2}(P_1^2+P_2^2)+\frac{\Omega_1}{2} Q_1^2+\frac{\Omega_2}{2} Q_2^2
    - 2 N \left(Q_1 Q_2^2 + 2 Q_1^3\right)
     + \frac{B}{2 Q_2^2},
}
\\
\displaystyle{
t=\varepsilon^{\et} T,\
q_1=\varepsilon^{\ea} \left(1 + \varepsilon Q_1\right),\
q_2=\varepsilon^{1+\ea} Q_2,\
n=\varepsilon^{-1-2 \ea-2 \et} N,\
}
\\
\displaystyle{
\alpha=\varepsilon^{4 \ea-2 \et-1}
 \left(N  - \frac{\Omega_1}{12} \varepsilon \right),\
\beta=\varepsilon^{4 \ea-2 \et+4} B,\
h=\varepsilon^{2 \ea -2 \et -1}
 \left(\frac{3}{2}N+\frac{\Omega_1}{4} \varepsilon + H \varepsilon^3\right),\
}
\\
\displaystyle{
\omega_1=\varepsilon^{-2 \et-1} \left(3 N + \frac{\Omega_1}{4} \varepsilon
 \right),\
\omega_2=\varepsilon^{-2 \et-1} \left(2 N + \Omega_2 \varepsilon \right).
% \varepsilon \to 0.
}
\end{array}
\right.
\label{eqConfluence_1:2:1_to_KdV5}
\end{eqnarray}

Since the three HH3 cases have been generated from some quartic case,
it is useless to find the general solution of the cubic cases.

It is quite instructive to also perform the confluence starting from
HH4 1:6:8.
In fact, there exist two mutually exclusive subcases
of 1:6:8 which are Liouville-integrable \cite{H1987}, these are
\begin{eqnarray}
\hbox{1:6:8a}
& &
H=(\ref{eqHH40}),\ \alpha=0,
\label{eqHH4_1:6:8a}
\\
\hbox{1:6:8b}
& &
H=(\ref{eqHH40})+\frac{\nu}{q_2^6},\ \gamma=0,
\label{eqHH4_1:6:8b}
\end{eqnarray}
and,
if one requires the presence of the inverse square term $q_1^{-2}$ in
the resulting HH3 case,
only the subcase HH4 1:6:8b is able
to achieve a confluence to a cubic case,
and only two cubic cases can be produced:
HH3 KK with an additional term $\nu_{\rm KK} q_2^{-6}$
(therefore also Liouville integrable \cite{H1987})
and HH3 KdV5 provided $\nu$ is nonzero.
With the definition
\begin{eqnarray}
& &
\left\lbrace
\begin{array}{ll}
\displaystyle{
h_{1:6:8b}(t) =
\frac{1}{2}(p_1^2+p_2^2) + \frac{\omega}{2} (4 q_1^2+ q_2^2)
 - \frac{n}{16} (8 q_1^4+ 6 q_1^2 q_2^2 + q_2^4)
}
\\
\displaystyle{
 +\frac{1}{2}\left(\frac{\alpha}{q_1^2}+\frac{\beta}{q_2^2}\right)
+\frac{\nu}{q_2^6},
}
\end{array}
\right.
\end{eqnarray}
the results are
\begin{eqnarray}
& &
\hbox{1:6:8b}\to\hbox{ KK}
\left\lbrace
\begin{array}{ll}
\displaystyle{
t=\varepsilon^{\et} T,\
q_1=\varepsilon^{\ea} \left(1 + \varepsilon Q_1\right),\
q_2=\varepsilon^{\ea+1} Q_2,\
n=\varepsilon^{-2 \ea-2 \et -1} N,\
}
\\
\displaystyle{
\alpha=\varepsilon^{4 \ea-2 \et-1}
       \left(-\frac{4}{3} N + 4 \Omega \varepsilon \right),\
\beta=\varepsilon^{4 \ea-2 \et+1} B,\
\nu=\varepsilon^{8 \ea-2 \et+8} \nu_{\rm KK},\
}
\\
\displaystyle{
\omega=\varepsilon^{-2 \et-1}
        \left(- N + \Omega \varepsilon \right),\
h=\varepsilon^{2 \ea -2 \et -1}
 \left(-2 N+4\Omega \varepsilon + H \varepsilon^2\right),\
}
\end{array}
\right.
\end{eqnarray}
and
\begin{eqnarray}
& &
\hbox{1:6:8b}\to\hbox{ KdV5}
\left\lbrace
\begin{array}{ll}
\displaystyle{
t=\varepsilon^{\et} T,\
q_1=\varepsilon^{\ea} Q_2,\
q_2=\varepsilon^{\ea-1} \left(1 + \varepsilon Q_1\right),\
n=\varepsilon^{-2 \ea-2 \et+1} N,\
}
\\
\displaystyle{
\alpha=\varepsilon^{4 \ea-2 \et} B,\
\beta=\varepsilon^{4 \ea-2 \et-5}
  \left(-\frac{N}{4}+\frac{1}{4}(2 \Omega_2-\Omega_1)\varepsilon \right),\
}
\\
\displaystyle{
\nu=\varepsilon^{8 \ea-2 \et-9}
  \left(\frac{N}{32}+\frac{1}{24}(\Omega_1-\Omega_2)\varepsilon \right),\
}
\\
\displaystyle{
\omega=\varepsilon^{-2 \et-1}
        \left(\frac{3 N}{16} + \frac{\Omega_2}{4} \varepsilon \right),\
h=\varepsilon^{2 \ea -2 \et -3}
 \left(-\frac{N}{16} + \frac{1}{12}(4 \Omega_2-\Omega_1)\varepsilon
 + H \varepsilon^3\right).
}
\end{array}
\right.
\end{eqnarray}

% ============================================================================
\section     {Integration of HH4 1:2:1 with a point transformation}
\label{sectionPoint}
\indent

The two cases HH4 1:2:1 and HH3 KdV5 are related by the
(one-way) confluence (\ref{eqConfluence_1:2:1_to_KdV5}),
but their relation is even stronger,
and there exists a point transformation
\cite[Eq.~(7.14)]{CMVCalogero}
between this quartic 1:2:1 case $H(Q_j,P_j,\Omega_1,\Omega_2,A,B)$
and the cubic KdV5 case $H(q_j,p_j,\omega_1,\omega_2,\alpha,\gamma)$.
\begin{eqnarray}
{\hskip -15.0 truemm}
& &
\left\lbrace
\begin{array}{ll}
\displaystyle{
Q_1^2 + Q_2^2 + \frac{\Omega_1+\Omega_2}{5}
= \alpha q_1  + \frac{\omega_1+4\omega_2}{20},\
}
\\
\displaystyle{
(\Omega_1-\Omega_2)(Q_1^2 - Q_2^2) =
\frac{\alpha^2}{2} q_2^2
- \frac{4 \omega_1 + 26 \omega_2}{5} \alpha q_1
- \frac{(\omega_1+4 \omega_2)^2}{100}
+2 E,
}
\\
\displaystyle{
\Omega_1=  \omega_1,\
\Omega_2=4 \omega_2,\
}
\end{array}
\right.
\label{eqFrom_121_to_KdV5}
\end{eqnarray}
Its action on the genus two hyperelliptic curve which integrates KdV5
\cite{Drach1919KdV} is just a translation.

It is worthwhile to notice that
the variables $Q_1$ and $Q_2$ of 1:2:1
and the variable $q_2$ of KdV5 are generically multivalued.

An attempt to find point transformations between the other quartic cases
and any cubic case has been unsuccessful for the moment.

% ============================================================================
\section     {Integration of the 1:6:1, 1:6:8, 1:12:16 cases
with birational transformations}
\label{sectionBirational}
\indent

The classification of fourth order first degree ODEs in the class
\begin{eqnarray}
& &
y''''=P(y'',y',y;t),
\label{eqclass_order4_deg1_poly}
\end{eqnarray}
in which $P$ is polynomial in $y'',y',y$ and analytic in $t$,
has been recently completed \cite{Cos2000a,Cos2000c}.
Although this class is too restrictive to include our equation
(\ref{eq_q1_Order4_form2}),
there exist transformations
\cite{CMVCalogero}
mapping
each HH4 case to at least one time-independent equation with the PP
in the class (\ref{eqclass_order4_deg1_poly}).
These transformations, which are birational transformations,
conserve the PP,
therefore they establish the PP for all the quartic cases
and, by confluence, for all the cubic cases.
Their explicit form is not very compact,
so we refer to Ref.~\cite{CMVCalogero} for further details.

% ==========================================================================
\section{Conclusion}

Although we have not yet finished to revisit the derivation of all
the two degree of freedom time-independent Hamiltonians
with the Painlev\'e property,
we thought it worthwhile to perform it starting from the basic principles,
so as to avoid any \textit{a priori} restriction on $V(q_1,q_2)$.

About the integration of the seven H\'enon-Heiles Hamiltonians,
the result, already summarized elsewhere \cite{CMVCalogero},
is the following.
All these seven Hamiltonians have a meromorphic general solution,
expressed with hyperelliptic functions of genus two,
therefore they have the Painlev\'e property.
Moreover,
these seven Hamiltonians are complete in the Painlev\'e sense,
i.e.~it is impossible to add any time-independent
term to the Hamiltonian without ruining
the Painlev\'e property.

% ==========================================================================
\section*{Acknowledgments}

The authors acknowledge the financial support of
the Tournesol grant no.~T2003.09.
RC warmly thanks the organizers for invitation.
CV is a postdoctoral fellow at the FWO-Vlaanderen.

% ***************************************************************** References

\label{lastpage}
\vfill \eject
\end{document}